\documentclass[11pt]{amsart} 
\usepackage{amscd,amssymb,amsxtra}
\usepackage{mathrsfs}  
\DeclareSymbolFontAlphabet{\mathrsfs}{rsfs}
\usepackage[mathscr]{eucal} 
\usepackage{mathabx}\DeclareMathSymbol{\emptyset}{\mathord}{symbols}{59}
                    \DeclareMathSymbol{\boxtimes}{\mathbin}{AMSa}{"02}
\usepackage{comment}
\usepackage{color}
\usepackage{enumitem} 
\usepackage{bbm}
\usepackage[colorlinks,backref=page,citecolor=refkey]{hyperref}
\usepackage{aliascnt}

\usepackage{marginnote}

\makeatletter
\let\@secnumfont\bfseries
\def\section{\@startsection{section}{1}%
  \z@{4\linespacing\@plus\linespacing}{\linespacing}%
  {\bfseries\centering}}
\def\introsection{\@startsection{section}{1}%
  \z@{3\linespacing\@plus\linespacing}{\linespacing}%
  {\bfseries\centering}}
\def\subsection{\@startsection{subsection}{2}%
   \z@{1.25\linespacing\@plus.7\linespacing}{.5\linespacing}%
   {\normalfont\bfseries}}
\def\subsectionsinline{\def\subsection{\@startsection{subsection}{2}%
  \z@{1\linespacing\@plus.7\linespacing}{-.5em}%
  {\normalfont\bfseries}}}

\makeatother

\numberwithin{equation}{section}

\newcommand{\mynewtheorem}[2]{
  \newaliascnt{#1}{equation}
  \newtheorem{#1}[#1]{#2}
  \aliascntresetthe{#1}
  \expandafter\def\csname #1autorefname\endcsname{#2}
}

\theoremstyle{definition}
\mynewtheorem{definition}{Definition}
\mynewtheorem{example}{Example}
\mynewtheorem{problem}{\color{blue}Problem}
\mynewtheorem{probsec}{\color{blue}Problem}
\mynewtheorem{exercise}{Exercise}
\mynewtheorem{question}{\color{blue}Question}
\mynewtheorem{project}{\color{blue}Project}
\mynewtheorem{construction}{Construction}
\mynewtheorem{task}{\color{blue}Task}
\mynewtheorem{otask}{\color{green}Obsolete Task}
\mynewtheorem{notation}{Notation}

\newtheorem*{definition*}{Definition}
\newtheorem*{example*}{Example}
\newtheorem*{problem*}{\color{blue}Problem}
\newtheorem*{probsec*}{\color{blue}Problem}
\newtheorem*{exercise*}{Exercise}
\newtheorem*{question*}{\color{blue}Question}
\newtheorem*{project*}{\color{blue}Project}
\newtheorem*{construction*}{Construction}
\newtheorem*{notation*}{Notation}

\theoremstyle{remark}
\mynewtheorem{note}{Note}
\mynewtheorem{remark}{Remark}
\mynewtheorem{data}{Data}

\newtheorem*{note*}{Note}
\newtheorem*{remark*}{Remark}
\newtheorem*{data*}{Data}

\theoremstyle{plain}
\mynewtheorem{theorem}{Theorem}
\mynewtheorem{corollary}{Corollary}
\mynewtheorem{lemma}{Lemma}
\mynewtheorem{proposition}{Proposition}
\mynewtheorem{conjecture}{Conjecture}
\mynewtheorem{claim}{Claim}
\mynewtheorem{proposal}{Proposal}
\mynewtheorem{conclusion}{Conclusion}
\mynewtheorem{hypothesis}{Hypothesis}
\mynewtheorem{assumption}{Assumption}

\newtheorem*{theorem*}{Theorem}
\newtheorem*{corollary*}{Corollary}
\newtheorem*{lemma*}{Lemma}
\newtheorem*{proposition*}{Proposition}
\newtheorem*{conjecture*}{Conjecture}
\newtheorem*{claim*}{Claim}
\newtheorem*{proposal*}{Proposal}
\newtheorem*{conclusion*}{Conclusion}
\newtheorem*{hypothesis*}{Hypothesis}
\newtheorem*{assumption*}{Assumption}

\newenvironment{proof*}[1][\proofname]{
  \begin{proof}[#1]}{  
\end{proof}}

\definecolor{mygreen}{rgb}{0.0, 0.85, 0.2}
\definecolor{orange}{rgb}{1.0, 0.55, 0.0}
\definecolor{refkey}{rgb}{0,.6,.4}

\renewcommand{\:}{\colon}

\DeclareMathOperator{\Aut}{Aut}
\newcommand{\CC}{{\mathbb C}}
\newcommand{\CP}{{\mathbb C\mathbb P}}

\DeclareMathOperator{\End}{End}

\newcommand{\HH}{{\mathbb H}}
\DeclareMathOperator{\Hom}{Hom}
\DeclareMathOperator{\id}{id}

\newcommand{\PP}{{\mathbb P}}

\newcommand{\RR}{{\mathbb R}}
\newcommand{\TT}{\mathbb T}
\DeclareMathOperator{\Spin}{Spin}

\newcommand{\ZZ}{{\mathbb Z}}

\newcommand{\chiup}{\raise.5ex\hbox{$\chi$}}

\newcommand{\inv}{^{-1}}
\DeclareRobustCommand{\mstrut}{^{\vphantom{1*\prime y\vee M}}}

\newcommand{\temsquare}{\raise3.5pt\hbox{\boxed{ }}}

\newcommand{\zmod}[1]{\ZZ/#1\ZZ}

\newcommand{\zt}{\zmod2}

\newcommand{\hneg}{\mkern-.5\thinmuskip}
 
\DeclareFontFamily{U}{mathx}{}
\DeclareFontShape{U}{mathx}{m}{n}{<-> mathx10}{}
\DeclareSymbolFont{mathx}{U}{mathx}{m}{n}
\DeclareMathAccent{\widehat}{0}{mathx}{"70}
\DeclareMathAccent{\widecheck}{0}{mathx}{"71}
 
\DeclareMathSymbol{\bigtimes}{1}{mathx}{"91}

\DeclareMathOperator{\SO}{SO}

\let\O\relax
\DeclareMathOperator{\O}{O}
\DeclareMathOperator{\U}{U}

\DeclareMathOperator{\GL}{GL}
\DeclareMathOperator{\SL}{SL}

\DeclareMathOperator{\PGL}{PGL}

\DeclareMathOperator{\PU}{PU}

\usepackage[all,2cell,color]{xy}
\UseAllTwocells
\usepackage{graphicx}\usepackage{epsf}

\DeclareMathOperator{\Cliff}{Cliff}
\DeclareMathOperator{\PQ}{PQ}
\DeclareMathOperator{\Pfaff}{Pfaff}
\DeclareMathOperator{\Q}{Q}
\newcommand{\AP}{\ssA_{\PP}}
\newcommand{\BLine}{B\kern .05em \Line}
\newcommand{\Bn}{\Bord_{\langle n-1,n  \rangle}}
\newcommand{\Bord}{\mathbf{Bord}}
\newcommand{\Btt}{\Bord_{\langle 2,3  \rangle}}
\newcommand{\Cx}{\CC^{\times }}

\newcommand{\IZ}{I\ZZ}
\newcommand{\Line}{\mathbf{Line}}
\newcommand{\Alg}{\mathbf{Alg}}
\newcommand{\MM}{\mathbb{M}}
\newcommand{\MSpin}{M\!\Spin}
\newcommand{\Man}{\mathbf{Man}}
\newcommand{\PH}{\PP\sH}
\newcommand{\POC}{\PP\Omega \sC}
\newcommand{\Proj}{\mathbf{Proj}}

\newcommand{\Vect}{\mathbf{Vect}}
\newcommand{\XT}{X\mstrut _T}
\newcommand{\aSG}{\alpha _{(\SS,\Gamma )}}
\newcommand{\bF}{\overline{F}}
\newcommand{\bG}{\overline{G}}
\newcommand{\brh}{\bar\rho }

\newcommand{\evo}[2]{e^{-i(t_{#1}-t_{#2})H/\hbar}}
\newcommand{\gpd}{/\!/} 
\newcommand{\hX}{\widehat{X}}
\newcommand{\hY}{\widehat{Y}}
\newcommand{\op}{^{\textnormal{op}}}
\newcommand{\sA}{\mathscr{A}}

\newcommand{\sCx}{\sC^{\times }}
\newcommand{\sC}{\mathscr{C}}
\newcommand{\sF}{\mathscr{F}}
\newcommand{\sG}{\mathscr{G}}
\newcommand{\sH}{\mathscr{H}}
\newcommand{\sL}{\mathcal{L}}
\newcommand{\sSet}{\mathbf{sSet}}

\newcommand{\ssA}{\mathrsfs{A}}
\newcommand{\tF}{\widetilde{F}}
\newcommand{\tG}{\widetilde{G}}
\newcommand{\ta}{\tilde\alpha }
\renewcommand{\SS}{\mathbb{S}}

\newcommand{\lact}{\;\reflectbox{\rotatebox[origin=c]{+90}{$\circlearrowleft$}}\;}

\DeclareTextFontCommand{\textbfit}{\bfseriesitshape}

\begin{document}

\abovedisplayskip18pt plus4.5pt minus9pt
\belowdisplayskip \abovedisplayskip
\abovedisplayshortskip0pt plus4.5pt
\belowdisplayshortskip10.5pt plus4.5pt minus6pt
\baselineskip=15 truept
\marginparwidth=55pt

\theoremstyle{definition}
\mynewtheorem{axiom}{Axiom System}

\makeatletter
\DeclareRobustCommand\bfseriesitshape{%
  \not@math@alphabet\itshapebfseries\relax
  \fontseries\bfdefault
  \fontshape\itdefault
  \selectfont
}
\renewcommand{\tocsection}[3]{%
  \indentlabel{\@ifempty{#2}{\hskip1.5em}{\ignorespaces#1 #2.\;\;}}#3}
\renewcommand{\tocsubsection}[3]{%
  \indentlabel{\@ifempty{#2}{\hskip 2.5em}{\hskip 2.5em\ignorespaces#1%
    #2.\;\;}}#3} 
\renewcommand{\tocsubsubsection}[3]{%
  \indentlabel{\@ifempty{#2}{\hskip 5.5em}{\hskip 5.5em\ignorespaces#1%
    #2.\;\;}}#3} 
\renewcommand\subsubsection{\@startsection{subsubsection}{3}%
  \z@{.5\linespacing\@plus.7\linespacing}{-.5em}%
  {\normalfont\bfseriesitshape}}
\def\@makefnmark{%
  \leavevmode
  \raise.9ex\hbox{\fontsize\sf@size\z@\normalfont\tiny\@thefnmark}} 
\def\multfoot{\textsuperscript{\tiny\color{red},}}
\def\footref#1{$\textsuperscript{\tiny\ref{#1}}$}
\makeatother

\setcounter{tocdepth}{2}

 \title[What is an anomaly?]{What is an anomaly?}
 \author[D. S. Freed]{Daniel S.~Freed}
 \thanks{\\[-6pt]This material is based on work supported by the National
Science Foundation under Grant Number DMS-2005286 and by the Simons Foundation
Award 888988 as part of the Simons Collaboration on Global Categorical
Symmetries.\\[3pt] I thank my longtime collaborators Mike Hopkins, Greg Moore,
Constantin Teleman and many others for influential conversations about
anomalies, which have taken place over a long period.  I also thank Theo
Johnson-Freyd for detailed feedback on an earlier draft.\\[-6pt]}
 \address{Department of Mathematics \\ Harvard University \\ Cambridge, MA
02138} 
 \email{dafr@math.harvard.edu}
 \date{July 14, 2023}
 \begin{abstract} 
 The anomaly of a quantum field theory is an expression of its projective
nature.  This starting point quickly leads to its manifestation as a special
kind of field theory: a once-categorified invertible theory.  We arrive at this
statement through a general discussion of projectivity and a discussion of
projectivity in quantum mechanics.  We conclude with a general formula for the
anomaly of a free spinor field.
 \end{abstract}
\maketitle

{\small
\def\reftext{References}
\renewcommand{\tocsection}[3]{%
  \begingroup 
   \def\tmp{#3}%
   \ifx\tmp\reftext
  \indentlabel{\phantom{QQQQ}\;\;} #3%
  \else\indentlabel{\ignorespaces#1 #2.\;\;}#3%
  \fi\endgroup}
\tableofcontents
}

This note\footnote{This is a written account of online talks given in February,
2023 in Arthur Jaffe's \emph{Picture Language} seminar and in Roger Picken's
\emph{TQFT Club} seminar.} is based on the thesis:
  \begin{equation}\label{eq:1}
     \textsc{Quantum theory is projective.  Quantization is linear.} \tag{A}
  \end{equation}
Hence our answer to the titular question:  
  \begin{equation}\label{eq:2}
     \textsc{The anomaly is the projectivity of a quantum theory.} \tag{B}
  \end{equation}
As such, the anomaly is a feature, not a bug~\cite{tH}.  It is an obstruction
only when quantizing~(\S\ref{sec:5}).
 
Anomalies in quantum field theory have a long history, dating back to work of
Steinberger~\cite{St}, Adler~\cite{Ad}, and Bell-Jackiw~\cite{BJ};
see~\cite{AgVm} for a recent survey and many additional references.  (We do not
attempt an account of that history here.)  The idea that an anomaly enjoys the
locality properties of a field theory was not explicit in the literature of the
20th century, at least to my knowledge, though the ``inflow'' mechanism of
Callen-Harvey~\cite{CH} is a precursor.  The first indications for me came from
reading~\cite{W1}.  This and many more influences led to~\cite{FT,F1}, in which
anomalies appear as invertible field theories.  (The first brief mention of
this formulation is at the end of~\cite{FHT}.)  Nonetheless, this previous work
does not tie anomalies in quantum field theory to projectivity, which is the
main point of this note.  Significantly, the new viewpoint in terms of
projective field theories leads us to define an anomaly as a
\emph{once-categorified invertible field theory}.

\subsubsection*{Two myths about anomalies}

First, it is often said that anomalies are only associated to symmetries.
However, the theory of a free spinor field (and no other fields) has an
anomaly.  For example, an anomaly is present in the minimal 1-dimensional
theory with a single spinor field, and there is no obvious symmetry to which to
ascribe the anomaly.  Second, one might think that anomalies are only caused by
fermionic fields.  Not so!  For example, the flavor symmetry of QCD is
anomalous---indeed, that anomaly involves fermions---but the anomaly persists
in the effective theory of pions, which is a bosonic theory in the sense that
there are no fluctuating fermionic fields.

  \section{Projective spaces, linearization, and symmetry} \label{sec:1}

We review the interplay between linear and projective geometry, emphasizing the
structure of symmetries.  The key concept is the \emph{projectivity} of a
projective representation.  We conclude with a discussion of the question:
\emph{What is projective space?}

\subsubsection*{Projectivization of a linear space}

Let $W$~be a vector space.  For the application to quantum theory we work over
the field of complex numbers, but the exposition generalizes to arbitrary
fields.  For simplicity, assume that $W$~is finite dimensional.  Associated
to~$W$ are two objects of a ``projective nature''.  The projective
space~$\PP(W)$ is the space of lines $L\subset W$.  (A \emph{line} is a
1-dimensional linear subspace.)  The algebra~$\End(W)$ is the space of linear
maps $T\:W\to W$.  One expression of the projective nature of~$\PP(W)$
and~$\End(W)$ is that they do not change when $W$~is replaced by the tensor
product~$W\otimes K$ with a line~$K$.  Namely, there are \emph{canonical}
isomorphisms
  \begin{equation}\label{eq:3}
     \begin{aligned}  \PP(W)&\longrightarrow \PP(W\otimes K) \\ 
            L\;\;\;&\longmapsto  \;\;\;\;\;L\otimes
            K\end{aligned}\qquad\qquad \qquad\begin{aligned} 
            \End(W)&\longrightarrow \End(W\otimes K) \\ 
            T\;\;&\longmapsto  \quad\;\;\;\;\;T\otimes
            \id_K\end{aligned}
  \end{equation}
These isomorphisms suggest that the algebra~$\End(W)$ is canonically associated
to the projective space~$\PP(W)$ in place of the vector space~$W$, a notion we
explain in footnote~\footref{assoc} below.

\subsubsection*{Projective symmetries}

A linear symmetry of~$W$ induces a projective symmetry of~$\PP (W)$, and
conversely a projective symmetry of~$\PP(W)$ has a $\Cx$-torsor\footnote{A
$\Cx$-torsor is a manifold equipped with a simply transitive action of~$\Cx$.
Any projective symmetry lifts to a linear symmetry, unique up to composition
with a \emph{homothety}: scaling by a nonzero complex number~$\lambda \in
\Cx$.} of lifts to a linear symmetry of~$W$.  The relationship between linear
and projective symmetries is encoded in the short exact sequence of Lie groups
  \begin{equation}\label{eq:4}
     \begin{gathered} \xymatrix{\Cx\ar[r] & \GL\ar[r] & \PGL & \phantom{B\Cx}}
      \end{gathered} 
  \end{equation}
which is a central extension.  We use notation for the finite dimensional model
space~$W=\CC^{n+1}$, though we drop the dimension~$n$; with appropriate
topologies these can also be taken to be groups of symmetries of infinite
dimensional spaces.  A Lie group~$G$ of \emph{projective} symmetries is
specified by a Lie group homomorphism $G\to \PGL$.  It induces a pull back
central extension of~$G$
  \begin{equation}\label{eq:5}
     \begin{gathered} \xymatrix{\Cx\ar[r] & \GL\ar[r] & \PGL&\phantom{B\Cx}\\
      \Cx\ar[r]\ar@<.5ex>@{=}[u]&\tG\ar[r]\ar@{-->}[u]&
      G\ar[u]} \end{gathered} 
  \end{equation}
together with a \emph{linear} action of the centrally extended group~$\tG$.   
 
A \emph{linearization} of the projective action is a lift of $G\to \PGL$ to a
homomorphism $G\to \GL$.  A linearization is equivalent to a \emph{splitting}
of the induced central extension:
  \begin{equation}\label{eq:6}
     \begin{gathered} \xymatrix{\Cx\ar[r] & \GL\ar[r] & \PGL &
     \phantom{B\Cx} \\ 
      \Cx\ar[r]\ar@<.5ex>@{=}[u]&\tG\ar[r]\ar@{-->}[u]&
      G\ar[u]\ar@[mygreen][ul]\ar@[mygreen]@<-1.1ex>[l]
     } \end{gathered}  
  \end{equation}
The \emph{obstruction} to a linearization is measured by a homomorphism~$\alpha
$ from~$G$ to a classifying space for the center~$\Cx$:
  \begin{equation}\label{eq:7}
     \begin{gathered} \xymatrix{\Cx\ar[r] & \GL\ar[r] & \PGL \ar[r]& B\Cx \\
      \Cx\ar[r]\ar@<.5ex>@{=}[u]&\tG\ar[r]^\pi \ar@{-->}[u]&
      G\ar[u]\ar@[red][ur]^<<<<<<{\alpha }}
      \end{gathered} 
  \end{equation}
One geometric model for~$B\Cx$ is the category of $\Cx$-torsors and
isomorphisms of $\Cx$-torsors.  The category~$B\Cx$ is a groupoid---all
morphisms are invertible---and furthermore it carries a symmetric tensor
product.  It is in this sense that we have a homomorphism $\alpha \:G\to B\Cx$:
for each~$g\in G$ there is a $\Cx$-torsor $L_g$; for each pair $g_1,g_2\in G$
there is an isomorphism $L_{g_1}\otimes L_{g_2}\to L_{g_1g_2}$; and for each
triple $g_1,g_2,g_3\in G$ there is a compatibility (associativity, cocycle
condition) among these isomorphisms.  In short, a homomorphism $G\to B\Cx$ is a
central extension of~$G$ by~$\Cx$.  The torsor~$L_g$ is $\pi \inv (g)$ in the
central extension.  The homomorphism $\alpha \:G\to B\Cx$ is the
\emph{projectivity} of the projective representation; it measures the deviation
from linearizability.
 
One can turn the story around.  Begin with the projectivity ~$\alpha $:
  \begin{equation}\label{eq:8}
     \begin{gathered} \xymatrix{\phantom{\Cx} & \phantom{\GL} & \phantom{\PGL
     }& B\Cx\\ 
      \phantom{\Cx}&\phantom{\tG}& G\ar@[red][ur]^<<<<<<{\alpha }}
      \end{gathered} 
  \end{equation}
and then define a projective representation with projectivity~$\alpha $ to be a
homomorphism~$\brh\:G\to \PGL$ that makes the diagram 
  \begin{equation}\label{eq:9}
     \begin{gathered} \xymatrix{\phantom{\Cx} & \phantom{\GL} & \PGL\ar[r]
     & B\Cx\\ \phantom{\Cx}&\phantom{\tG}&
     G\ar[u]^{\brh}\ar@[red][ur]^<<<<<<{\alpha }} 
     \end{gathered} 
  \end{equation}
commute.  As argued above, the projectivity~$\alpha $ is equivalent to a
central extension of~$G$ by~$\Cx$, and then a projective representation~$\brh$
of~$G$ is equivalent to a linear representation~$\rho $ of~$\tG$ in which the
central~$\Cx$ acts by scalar multiplication:
  \begin{equation}\label{eq:10}
     \begin{gathered} \xymatrix{\Cx\ar[r] & \GL\ar[r] & \PGL\ar[r] &
     B\Cx\\ {\Cx}\ar@<.5ex>@{=}[u]\ar[r]&{\tG}\ar[r]\ar[u]^{\rho }&
     G\ar[u]^{\brh}\ar@[red][ur]^<<<<<<{\alpha 
     }} \end{gathered} 
  \end{equation}
A linearization of~$\brh$ is equivalent to a splitting of the central
extension, which in other terms is a \emph{trivialization} of the
projectivity~$\alpha $: 
  \begin{equation}\label{eq:11}
     \begin{gathered} \xymatrix{\Cx\ar[r] & \GL\ar[r] & \PGL\ar[r] & B\Cx\\
     {\Cx}\ar@<.5ex>@{=}[u]\ar[r]&{\tG}\ar[r]\ar[u]^{\rho }&
     G\ar[u]^{\brh}\ar@[mygreen][ul]\ar@[mygreen]@<-1.1ex>[l]
     \urlowertwocell<-6>_1{^{\color{mygreen}\simeq}\;\; }\ar@[red][ur]^<<<<<<{\alpha }
     } \end{gathered}  
  \end{equation}

\subsubsection*{Cohomological interpretation}

The \emph{homotopy class} or \emph{isomorphism class} (the appropriate phrase
depends on the model of~$B\Cx$) of the projectivity~$\alpha $ lies in a
cohomology group $H^2(G;\Cx)$.  The precise type of cohomology depends on the
nature of the group~$G$.  For example, if $G$~is a finite group, then we can
use standard group cohomology of Eilenberg-MacLane, which is equivalent to the
singular cohomology of a space that realizes the classifying space~$BG$.  If
$G$~is a Lie group, then a different type of cohomology is needed.  As
intimated several times now, the central extension
  \begin{equation}\label{eq:12}
     \begin{gathered} \xymatrix{\phantom{\Cx} & \phantom{\GL} & \phantom{\PGL}&
     B\Cx \\ 
      \Cx\ar[r]&\tG\ar[r]&
      G\ar@[red][ur]^<<<<<<{\alpha }} \end{gathered} 
  \end{equation}
is a ``cocycle'' for the cohomology class.  Splittings of the central
extension---equivalently trivializations of the projectivity~$\alpha $---form a
torsor over characters~$\lambda $ of~$G$:
  \begin{equation}\label{eq:13}
     \begin{gathered} \xymatrix{\phantom{\Cx} & \phantom{\GL} &
     \phantom{\PGL}& B\Cx \\ \Cx\ar[r]&\tG\ar[r]&
     G\urlowertwocell<-6>_1{^{\color{mygreen}\simeq}\;\; } 
     \ar@[mygreen]@<-1.1ex>[l]\ar@[red][ur]^<<<<<<{\alpha
     }\ar@/^1pc/@[orange][ll]^{\lambda }} \end{gathered}  
  \end{equation} 
Characters are \emph{invertible} linear representations, or in terms of
cohomology, elements of $H^1(G;\Cx)$.  Recall that the class of the
projectivity lies in the next higher cohomology group $H^2(G;\Cx)$.  This leads
to the main takeaway: 
  \begin{equation}\label{eq:15}
  \begin{gathered}
     \textsc{The projectivity of a projective representation is a ``suspended''
     or}\\[-6pt] \textsc{``delooped'' or ``once-categorified'' invertible linear
     representation.} 
  \end{gathered}
  \end{equation}

\subsubsection*{What is a projective space?}

In other words, how do we define a projective space without expressing it as
the projectivization~$\PP(W)$ of a linear space~$W$?  Our answer is a special
case of a definition of Klein-Cartan.  A \emph{model geometry} $H\lact X$
consists of a smooth manifold~$X$ equipped with a left action of a Lie
group~$H$.  For example, the model for $n$-dimensional complex linear geometry
is $\GL_n\!\CC\lact\CC^n$.  The model for $n$-dimensional hyperbolic geometry
is $\O_{n,1}\!\lact \HH^n$, where $\HH^n$~is hyperbolic space.  An instance of
$(H\lact X)$-geometry is specified by a right $H$-torsor~$T$; the associated
manifold
  \begin{equation}\label{eq:14}
     \XT := T\times _{H}X 
  \end{equation}
carries this geometry.  Here $T\times _{H}X $~is the mixing construction:
$T\times _{H}X =(T\times X)/H$, where $h\in H$ transforms $(t,x)\in T\times X$
by $(t,x)\cdot h = (t\cdot h,h\inv \cdot x)$.  (Example: for $n$-dimensional
complex linear geometry, $\XT$~is a complex vector space and $T$~is the right
$\GL_n\!\CC$-torsor of isomorphisms $\CC^n\xrightarrow{\;\cong \;}\XT$: bases
of~$\XT$.)  If $S$~is a smooth manifold, then a family of $(H\lact
X)$-geometries is specified by a principal $H$-bundle $P\to S$.  The geometry
lives on the associated fiber bundle $X\!\mstrut _P\to S$.  Furthermore, we
allow $S$~to be a smooth stack.  Example: if $S = *\gpd G$ for a Lie group~$G$,
then an $(H\lact X)$-geometry over~$S$ is a single $(H\lact X)$-geometry
equipped with a $G$-action.
 
With this in mind, the model for $n$-dimensional complex projective geometry
has $X=\CP^n$.  There are several possibilities for the group~$H$.  For complex
algebraic geometry $H=\PGL_{n+1}\!\CC$ is the group of all holomorphic
automorphisms of~$\CP^n$.  In K\"ahler geometry $H=\PU_{n+1}$ is the group of
holomorphic isometries of~$\CP^n$.  There are infinite dimensional analogs.  We
will see below that neither is the correct choice for the model geometry for
quantum mechanics.

  \section{Quantum mechanics as a projective system}\label{sec:2}

We begin with the usual linear description of quantum mechanics, and then point
out its projective nature.  Wigner's theorem is used to determine the correct
model geometry for quantum mechanics.

\subsubsection*{Quantum mechanics as a linear system}

Typically, one gives the \emph{state space}~$\sH$, assumed to be a complex
separable Hilbert space, together with a self-adjoint operator~$H$ on~$\sH$,
the Hamiltonian.  From this data one defines the space $\PH$ of pure states, a
convex space of all states, and an algebra of observables.  The space of pure
states comes equipped with a function
  \begin{equation}\label{eq:16}
     \begin{aligned} p\:\PH\times \PH&\longrightarrow \;\;[0,1] \\ L_0\;,\;
      L_1\;&\longmapsto |\langle \psi _0,\psi _1 \rangle|^2\qquad
      \end{aligned} 
  \end{equation}
that encodes transition probabilities between pure states.  Here $\psi _i\in
L_i$, $i=0,1$, are unit norm vectors in the lines $L_0,L_1\in \PH$. 
 
Suppose given times (real numbers) $t_0<t_1<\cdots< t_n<t_{f}$, initial and
final pure states $L _0,\;L _{f}\in \PH$, and observables\footnote{For us
`$\End(\sH)$' denotes \emph{bounded} operators, and for this formal discussion
we restrict to bounded observables.} $A_1,\dots ,A_n\in\End(\sH)$.  Then the
basic quantity in the theory is
  \begin{equation}\label{eq:17}
     p\hneg\left( L _{f}\,,\, \evo{f}n A_n\cdots\evo21 A_1\evo10\, L _0
     \right)\;\in[0,1] 
  \end{equation}
This probability is the norm square of a complex amplitude.  If we choose unit
vectors $\psi _0\in L_0$, $\psi _f\in L_f$, then the amplitude is the complex
number
  \begin{equation}\label{eq:18}
     \bigl\langle \psi _{f}\,,\, \evo{f}n A_n\cdots\evo21 A_1\evo10 \,\psi _0
     \bigr\rangle_{\sH} \;\in \CC 
  \end{equation}
Of course, we should write the amplitude in terms of the states~$L_0,L_f$, not
in terms of the auxiliary choices~$\psi _0,\psi _f$.  This is straightforward:
as a function of~$\psi _0,\psi _f$, \eqref{eq:18}~is an element of the
hermitian line $(\overline{L_f}\otimes L_0)^*\cong L_f\otimes \overline{L_0}$. 

It is striking that ordinary amplitudes in quantum mechanics are elements of
complex lines rather than complex numbers.

\subsubsection*{Quantum mechanics as a projective system}

As explained in~\S\ref{sec:1}, the space $\PH$ of pure states and the space
$\End(\sH)$ of observables only depend on the projective space that underlies
the linear space~$\sH$.  Therefore, suppose a projective space~$\PP$ in the
sense of~\eqref{eq:14} is given.  (Shortly we determine precisely what model
geometry we need.)  Then $\PP$~is the space of pure states of a quantum
mechanical system, and there is an associated algebra~$\AP$ of
observables.\footnote{\label{assoc}If $H\lact X$ is the model geometry, with
$X=\CP=\PP(\sH)$~a model complex projective space of finite or infinite
dimension, so $\sH=\CP^N$ or~$\sH$~is a standard infinite dimensional Hilbert
space; and if $T$~is the right $H$-torsor that specifies the projective space
$\PP = T\times _H\CP$; then $\AP = T\times _H\End(\sH)$, where $H$~acts by
conjugation on~$\End(\sH)$.}  The transition probability function
  \begin{equation}\label{eq:19}
     \begin{aligned} p\:\PP\times \PP\,&\longrightarrow \;\;\;[0,1] \\ \sigma
      _0\,,\, \sigma _1&\longmapsto |\langle \psi _0,\psi _1
      \rangle|^2_{\sH}\qquad \end{aligned} 
  \end{equation}
is defined by choosing a linearization $\PP\xrightarrow{\;\cong \;}\PH$ and
choosing unit norm vectors $\psi _0,\psi _1$ in the lines of~$\sH$ that
correspond to~$\sigma _0,\sigma _1$.  One checks\footnote{Use that $H=\PQ$, and
that a linearization of the projective space associated to a $\PQ$-torsor~$T$
is associated to a lift of~$T$ to a $Q$-torsor.  (The Lie groups~$\Q$ and~$\PQ$
are defined in the next subsection.)  The collection of lifts, and so of
linearizations, forms a \emph{$\TT$-gerbe}.  See footnote~\footref{gerbe} below
for one model of a gerbe.} that \eqref{eq:19}~is independent of these choices.

With this setup in place, the analog of the probability~\eqref{eq:17} is  
  \begin{equation}\label{eq:20}
     p\hneg\left( \sigma _{f}\,,\, \evo{f}n A_n\cdots\evo21 A_1\evo10\, \sigma _0
     \right)\;\in[0,1] 
  \end{equation}
The amplitude~\eqref{eq:18} is also well-defined in this projective setup; it
lives in a hermitian line $\sL_f\otimes \overline{\sL_0}$ that is defined
using~\eqref{eq:18} and the gerbe of linearizations of~$\PP$ .  As before, the
probability~\eqref{eq:20} is the norm square of this amplitude.

\subsubsection*{The symmetry/structure group of quantum mechanics}

This is the group~$\PQ$ of permutations of the points of a projective
space~$\PP$ that preserve the function~\eqref{eq:19}, which is the basic data
of quantum mechanics.  Fix a linearization $\PP\xrightarrow{\;\cong \;}\PH$;
then the group~$\PQ=\Aut(\PP,p)$ of maps~$\PP\longrightarrow \PP$
preserving~$p$ is the isometry group of the Fubini-Study metric ~$ d\:\PH\times
\PH\longrightarrow \RR^{\ge0}$.  This follows from the well-known formula
(see~\cite{BH} for a discussion and references)
  \begin{equation}\label{eq:21}
     \cos(d)=2p-1 
  \end{equation}
The structure group~$\PQ$ we seek is the isometry group of a model Fubini-Study
projective space.

  \begin{example}[]\label{thm:2}
 If $\sH=\CC^2$, then $\PP=\CP^1$ is the complex projective line, which is
diffeomorphic to the 2-sphere~$S^2$, and the Fubini-Study metric transports to
the round metric on~$S^2$.  The isometry group in this case is the orthogonal
group~$\O_3$.  Observe the group extensions 
  \begin{equation}\label{eq:22}
     \begin{aligned} \TT&\longrightarrow &&\!\!\U_2\longrightarrow
      &&\!\!\!\SO_3 \\[.3em] \TT&\longrightarrow &&\!\!\Q_2\longrightarrow
      &&\!\!\!\phantom{n}\O_3 =\PQ_2\end{aligned} 
  \end{equation}
with kernel the group $\TT\subset \Cx$ of unit norm complex numbers.  The first
is a central extension of the identity component~$\SO_3$ of the Lie group
$\PQ_2=\O_3$.  Elements of~$\SO_3$ act projectively on~$\CC^2$, and $\U_2$~ is
the group of lifts to isometries of~$\CC^2$.  The second extension
in~\eqref{eq:22} is not central.  The extended group~$\Q_2$ consists of unitary
and antiunitary automorphisms of~$\CC^2$, and antiunitary automorphisms
anticommute with elements of the kernel~$\TT$.
  \end{example}

The following result, known as \emph{Wigner's theorem}, is undoubtedly due to
Cartan for finite dimensional projective spaces (in the form of determining the
isometry group; see~\cite{Lo}), and in the infinite dimensional case appears
first in~\cite{vNW} without proof; see~\cite{Bo} for a history and~\cite{F2}
for two geometric proofs and references to earlier proofs.  Let $\PP$~be the
projectivization of~$\sH=\CC^N$ in the finite dimensional case and of a model
complex separable Hilbert space~$\sH$ in the infinite dimensional case.  Let
$\PQ$~denote the group of automorphisms of~$\PH$ that preserve the
function~\eqref{eq:19}.

  \begin{theorem}[von Neumann--Wigner]\label{thm:3}
 The group~$\PQ$ of projective quantum mechanical symmetries fits into a group
extension 
  \begin{equation}\label{eq:23}
     \TT\longrightarrow \Q\longrightarrow \PQ
  \end{equation}
where $\Q$ is the group of unitary and antiunitary transformations of~$\sH$.  
  \end{theorem}

\noindent
 The import of the theorem is the surjectivity of the map $\Q\to \PQ$.  Each
of~$\Q$ and~$\PQ$ is a Lie group with two components; the identity component
of~$\Q$ is the unitary group.  Let $\Q_N, \PQ_N$ be these Lie groups for
$\sH=\CC^N$, and let $\Q_\infty ,\PQ_\infty $ be these groups in the infinite
dimensional case.   
 
This discussion is summarized by the statement ($n\in \ZZ^{>0}\cup \{\infty
\}$):
  \begin{equation}\label{eq:24}
     \textsc{The model geometry for quantum mechanics is $\PQ_{n+1}\!\lact\,
     \CP^{n}$} 
  \end{equation}

  \section{Quantum field theory as a projective system}\label{sec:3}

Now we arrive at the central section of the paper.  We begin by reviewing the
Segal axioms for Wick-rotated field theory, in which a field theory is a linear
representation of a bordism category.  Then, following the main
theme~\eqref{eq:1}, we indicate the variation in which a field theory is a
projective representation of a bordism category.  The anomaly is the
projectivity of that representation, and our main conclusion---analogous
to~\eqref{eq:15}---is that the anomaly is a once-categorified invertible field
theory.  We argue that the physical content of a quantum field theory is
contained in a projective theory rather than a linear theory.  In most cases we
know, the anomaly extends to a one higher dimensional ordinary invertible field
theory, as we indicate at the end of this section.

\subsubsection*{Wick-rotated field theory as a linear representation}

Segal~\cite{S1} first introduced his axiom system in the context of
2-dimensional conformal field theory.  It was taken over to topological field
theory by Atiyah~\cite{A}, and subsequently extended in many directions.  Over
time it became clear that the axioms should apply to general quantum field
theories.  This general context is the subject of a recent paper by
Kontsevich-Segal~\cite{KS}.  For our purposes here, we emphasize the formal
structure. 
 
There are two discrete parameters that specify the ``type'' of field theory.
The first is a positive integer~$n$, which in Lorentz signature is the
dimension of spacetime.  The second is the collection~$\sF$ of background
fields.  (In this axiom system there are no fluctuating fields---they have
already been integrated out---but see~\S\ref{sec:5}.)  Let $\Man_n$ denote the
category of smooth $n$-manifolds and local diffeomorphisms, and let $\sSet$
denote the category of simplicial sets.

  \begin{definition}[]\label{thm:4}
 A \emph{Wick-rotated field} is a sheaf 
  \begin{equation}\label{eq:25}
     \sF\:\Man_n\op\longrightarrow \sSet 
  \end{equation}
  \end{definition}

\noindent
 One needs extra structure on~$\Man_n$ to formulate the sheaf condition, which
as usual is in terms of open covers; see~\cite{FH1} for one exposition.  Some
fields take values in the category of sets: Riemannian metrics, $\RR$-valued
functions, $N$-valued functions for a fixed smooth manifold~$N$, orientations,
etc.  Others take values in groupoids: $G$-connections for a fixed Lie
group~$G$, spin structures, etc.  Still others take values in higher groupoids:
$B$-fields, $C$-fields, etc.  The choice of $\sSet$ as codomain
in~\eqref{eq:25} accommodates all of these.  In our usage, $\sF$~is a finite
set of fields, such as $\sF=\{\textnormal{orientations, Riemannian metrics}\}$.
For an $n$-manifold~$M$, the space of sections~$\sF(M)$ of the sheaf is the
space (simplicial set) of fields on~$M$.

  \begin{figure}[ht]
  \centering
  \includegraphics[scale=1.2]{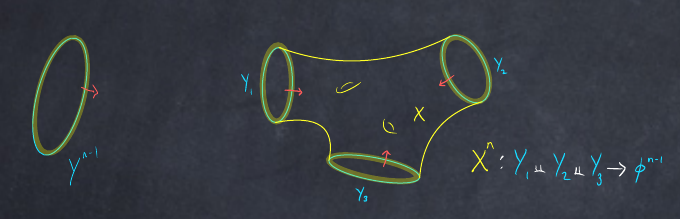}
  \vskip -.5pc
  \caption{An object and a morphism in the bordism category $\Bord_n(\sF)$.
  Germs of $n$-manifolds are represented by the shadings, and coorientations by
  the red arrows.  Background fields are not depicted.}\label{fig:1}
  \end{figure}

Given $n,\sF$, we define a bordism category $\Bord_{\langle n-1,n
\rangle}(\sF)$.  An object is a closed $(n-1)$-manifold~$Y$, embedded in a
germ~$\hY$ of an $n$-manifold, equipped with a coorientation in that germ, and
equipped with an object of~$\sF(\hY)$.  (The coorientation is the Wick-rotated
remnant of an ``arrow of time'' of a spacelike hypersurface.)  A morphism
$X\:Y\to Y'$ is a compact $n$-manifold with boundary equipped with a partition
into incoming and outgoing components, a germ~$\hX$ of an embedding of the
boundary into an $n$-manifold that extends the germ of a collar neighborhood of
the boundary, isomorphisms of the incoming boundary germs with~$\hY$ and the
outgoing boundary germs with~$\hY'$, and an object of~$\sF(\hX)$ together with
isomorphisms of its restrictions to the boundary germs with the background
fields on~$\hY$ and~$\hY'$.  A sample object and morphism are depicted in
\autoref{fig:1}.  Composition is by gluing, and disjoint union provides a
symmetric monoidal structure.

  \begin{remark}[]\label{thm:5}
 A field or collection of fields~\eqref{eq:25} is \emph{topological} if the
sheaf that defines it is \emph{locally constant}.  Examples include
orientations, spin structures, $G$-bundles for a finite group~$G$, etc.
Non-examples include $\RR$-valued functions, Riemannian metrics,
$G$-connections for a positive dimensional Lie group~$G$, etc.  A locally
constant sheaf is equivalent to an $n$-dimensional tangential structure in the
sense introduced into bordism theory by Lashof~\cite{La}.  For topological
fields the bordism category was first introduced by Milnor in his proof of the
h-cobordism theorem~\cite{M}.
  \end{remark}

Let $\Vect$ denote a suitable category of topological vector spaces and linear
maps; see \cite[\S3]{KS} for details.  The important point is that the
linear maps are \emph{nuclear}, which is a form of smallness.\footnote{In
quantum mechanics, for example, in real time the evolution is by unitary
operators~$e^{-itH/\hbar}$, $t\in \RR$, whereas Wick-rotated ``evolution'' is
by smoothing maps~$e^{-\tau H/\hbar}$, $\tau \in \RR^{>0}$, which in particular
are nuclear (trace class).}   

  \begin{axiom}[]\label{thm:6}
 A (Wick-rotated) \emph{field theory} 
  \begin{equation}\label{eq:26}
     F\:\Bord_{\langle n-1,n  \rangle}(\sF)\longrightarrow \Vect 
  \end{equation}
is a linear representation of a bordism category.
  \end{axiom}

\noindent
 Here `linear representation' means a symmetric monoidal functor of symmetric
monoidal categories, which we simply call a `homomorphism'.  So $F$~assigns a
vector space (state space)~$F(Y)$ to every closed $(n-1)$-manifold and a linear
map $F(X)\:F(Y)\to F(Y')$ to every bordism $X\:Y\to Y'$.

  \begin{remark}[]\label{thm:8}
 \ 
 \begin{enumerate}[label=\textnormal{(\arabic*)}]

 \item A field theory is \emph{topological} if it factors through a bordism
category $\Bord_{\langle n-1,n \rangle}(\sF')$ in which $\sF'$~is a locally
constant sheaf.

 \item For topological theories there is an extended notion of \emph{locality}.

 \item Just as representations of Lie groups are not necessarily unitarizable
(example: $\SL_2\!\RR$), so too the notion of \emph{unitarity} is not part of
\autoref{thm:6}.  Rather, unitarity appears in the form of reflection
positivity, which is a structure on~$F$ that we do not elaborate here.
 \end{enumerate}
  \end{remark}

\subsubsection*{Wick-rotated QFT as a projective representation}

As in the case of quantum mechanics~(\S\ref{sec:2}), quantum field theory is a
projective system, and so \eqref{eq:26} should take values in a suitable
category of projective spaces and projective linear maps.   

  \begin{remark}[]\label{thm:9}
 At the end of~\S\ref{sec:1} we gave several model geometries for complex
projective space.  Since unitarity is not built into \autoref{thm:6}, the
appropriate model for the projective variation of the axioms is that of complex
algebraic geometry, namely $\PGL_{n+1}\hneg\CC\,\lact \,\CP^{n}$, where $n\in
\ZZ^{>0}\cup \{\infty \}$.  (For $n=\infty $ we use `$\CP^{\infty}$' for the
projective space of a distinguished object of~$\Vect$, in the same sense that
$\CC^N$ is a distinguished $N$-dimensional complex vector space.)
  \end{remark}

Let $\Line$ be the category of 1-dimensional complex vector spaces and
invertible linear maps between them; it is a categorified version of the
group~$\Cx$.  We seek a categorification of the short exact
sequence~\eqref{eq:4}, which we write as:
  \begin{equation}\label{eq:27}
     \begin{gathered} \xymatrix@R-3.5pc{\Line \ar[r] & \Vect\ar[r] & \Proj
     & 
      \phantom{\BLine} \\ \phantom{\Line}& \phantom{\widetilde{\Bord_{\langle n-1,n  \rangle}}(\sF)}&
      \phantom{\Bord_{\langle n-1,n  \rangle}(\sF)}} \end{gathered} 
  \end{equation}
Heuristically, $\Proj$~is a category of complex projective spaces and
projective linear maps, and it should be based on the particular topological
vector spaces and continuous linear maps used to define the category~$\Vect$.
More precisely, the Picard groupoid~$\Line$, which is a model for~$B\Cx$, acts
on~$\Vect$ and $\Proj$ is the quotient of this action.  This quotient is
naturally a 2-category, not a 1-category.  See Appendix~\ref{sec:7} for more
indications of the definition of $\Proj$.  As in~\eqref{eq:7} there is a
classifying space for~$\Line$ and an extension of the sequence~\eqref{eq:27}:
  \begin{equation}\label{eq:28}
     \begin{gathered} \xymatrix@R-3.5pc{\Line \ar[r] & \Vect\ar[r] & \Proj
     \ar[r] & 
      \BLine \\ \phantom{\Line}& \phantom{\widetilde{\Bord_{\langle n-1,n  \rangle}}(\sF)}&
      \phantom{\Bord_{\langle n-1,n  \rangle}(\sF)}} \end{gathered} 
  \end{equation}
 
We have the ingredients in place to mimic the discussion in~\S\ref{sec:1} that
begins with~\eqref{eq:5}.  Namely, a \emph{projective field theory}~$\bF$ is a
homomorphism (symmetric monoidal functor) into~$\Proj$:
  \begin{equation}\label{eq:33}
     \begin{gathered} \xymatrix{\Line \ar[r] & \Vect\ar[r] & \Proj \ar[r] &
      \BLine \\ \phantom{\Line}& \phantom{\widetilde{\Bord_{\langle n-1,n  \rangle}}(\sF)}&
      \Bord_{\langle n-1,n  \rangle}(\sF)\ar[u]_{\bF}} \end{gathered} 
  \end{equation}
Its \emph{projectivity}, or \emph{anomaly}~$\alpha $, is the composition with
the map to $\BLine$:
  \begin{equation}\label{eq:32}
     \begin{gathered} \xymatrix{\Line \ar[r] & \Vect\ar[r] & \Proj \ar[r] &
      \BLine \\ \phantom{\Line}&
      \phantom{\widetilde{\Bord_{\langle n-1,n  \rangle}}(\sF)}&
      \Bord_{\langle n-1,n  \rangle}(\sF)\ar[u]_{\bF}\ar@[red][ur]^<<<<<<{\alpha }} \end{gathered} 
  \end{equation}
Since $\alpha $~is a map out of a bordism category, it is a field theory.
Shortly we identify its precise nature.  At this stage we simply emphasize the
analogy with the map~$\alpha $ in~\eqref{eq:7}, which measures the projectivity
of a projective representation of a Lie group.  Analogous to~\eqref{eq:6}, a
projective field theory induces a central extension of a bordism category and a
linear representation~$\tF$ of the centrally extended bordism category:
  \begin{equation}\label{eq:31}
     \begin{gathered} \xymatrix{\Line \ar[r] & \Vect\ar[r] & \Proj \ar[r] &
      \BLine \\ \Line\ar@{=}[u]\ar[r] &
      \widetilde{\Bord_{\langle n-1,n  \rangle}}(\sF)\ar@{-->}[u]_{\tF}\ar[r] &
      \Bord_{\langle n-1,n  \rangle}(\sF)\ar[u]_{\bF}\ar@[red][ur]^<<<<<<{\alpha }} \end{gathered} 
  \end{equation}
This version of a projective field theory---a linear representation of a
centrally extended bordism category---already appears in Segal's original
paper, namely \cite[Definition~(5.2)]{S1} (which in fact is more general).  A
trivialization of the anomaly is equivalent to a lift of the projective
theory~$\tF$ to a linear theory~$F$:
  \begin{equation}\label{eq:30}
     \begin{gathered} \xymatrix{\Line \ar[r] & \Vect\ar[r] & \Proj \ar[r] &
      \BLine \\ \Line\ar@{=}[u]\ar[r] &
      \widetilde{\Bord_{\langle n-1,n  \rangle}}(\sF)\ar@{-->}[u]_{\tF}\ar[r] &
      \Bord_{\langle n-1,n  \rangle}(\sF)\ar[u]_{\bF}\urlowertwocell<-6>_1{^{\color{mygreen}\simeq}\;\; } 
     \ar@[mygreen]@<-1.1ex>[l]\ar@[red][ur]^<<<<<<{\alpha
     } \ar@<-1.1ex>@[green][l] \ar@[green][ul]_<<<<<{F}} \end{gathered} 
  \end{equation}
Trivializations of~$\alpha $---if they exist---form a torsor over homomorphisms
$\lambda \:\Bord_{\langle n-1,n \rangle}(\sF)\to \Line$:
  \begin{equation}\label{eq:29}
  \begin{gathered} \xymatrix{\Line \ar[r] &
      \Vect\ar[r] & \Proj \ar[r] 
      &  \BLine \\
      \Line\ar@{=}[u]\ar[r]  &
  \widetilde{\Bord_{\langle n-1,n \rangle}}(\sF)\ar@{-->}[u]_{\tF}\ar[r] &
  \Bord_{\langle n-1,n
  \rangle}(\sF)\ar[u]_{\bF}\urlowertwocell<-6>_1{^{\color{mygreen}\simeq}\;\; }
     \ar@[mygreen]@<-1.1ex>[l]\ar@[red][ur]^<<<<<<{\alpha
     }  \ar@<-1.1ex>@[green][l]
  \ar@[green][ul]_<<<<<{F}\ar@/^1pc/@[orange][ll]^{\lambda }} 
  \end{gathered} 
  \end{equation}

\subsubsection*{Anomaly as a once-categorified invertible field theory}

We begin by elucidating the classifying space $\BLine$, which we model as a
2-category, just as $B\Cx$ is modeled as a 1-category after~\eqref{eq:7}.  A
simple model is the 2-groupoid~$B^2\Cx$ which has a single object, a single
1-morphism, and the group~$\Cx$ of 2-morphisms.  A more robust model is the
2-category of \emph{$\Cx$-gerbes},\footnote{\label{gerbe}Here is one model of a
$\Cx$-gerbe.  The data is a set~$\sA$; for each pair $\alpha _0,\alpha _1\in
\sA$ a complex line $L_{\alpha _0\alpha _1}$; and for each triple $\alpha
_0,\alpha _1,\alpha _2\in \sA$ an isomorphism of lines $\theta _{\alpha
_0,\alpha _1,\alpha _2}\:\CC\xrightarrow{\;\cong \;}L_{\alpha _1\alpha
_2}\mstrut \otimes L_{\alpha _0\alpha _2}\inv \otimes L_{\alpha _0\alpha _1}$.
For each quartet $\alpha _0,\alpha _1,\alpha _2,\alpha _3\in \sA$ there is a
condition on the~$\theta $'s.  A projective space~$\PP$ gives rise to a gerbe
in which $\sA$~is a set of linearizations of~$\PP$.} or equivalently of
\emph{invertible $\Vect$-modules}.

  \begin{definition}[]\label{thm:10}
 Fix $n,\sF$.
 \begin{enumerate}[label=\textnormal{(\arabic*)}]

 \item An \emph{invertible field theory} is a homomorphism 
  \begin{equation}\label{eq:34}
     \lambda \:\Bord_{\langle n-1,n  \rangle}(\sF) \longrightarrow \Line 
  \end{equation}

 \item A \emph{once-categorified invertible field theory} is a homomorphism
  \begin{equation}\label{eq:35}
     \quad\!\alpha  \:\Bord_{\langle n-1,n  \rangle}(\sF) \longrightarrow \BLine 
  \end{equation} 
 \end{enumerate} 
  \end{definition}

\noindent
 Let $Y$~be an object of~$\Bord_{\langle n-1,n \rangle}(\sF)$, which is
essentially a closed $(n-1)$-manifold.  (The additional data of an object---the
germ of a cooriented embedding into an $n$-manifold together with a background
field---is spelled out following \autoref{thm:4}.)  Then $\lambda (Y)$~is a
complex line, the 1-dimensional state space of the invertible field
theory~$\lambda $ on~$Y$.  By contrast, $\alpha (Y)$~is a gerbe.  If $\alpha
$~is the anomaly of a projective theory~$\bF$, then $\alpha (Y)$~measures the
projectivity of the projective space~$\bF(Y)$.  Now suppose $X$~is a closed
$n$-manifold equipped with a background field.  Then $\lambda (X)$~is a nonzero
complex number, whereas $\alpha (X)$~is a complex line.  If $\bF$~is a
projective theory with anomaly~$\alpha $, then $\bF(X)$ is an element of the
complex line~$\alpha (X)$.  (Recall that amplitudes in quantum mechanics are
naturally elements of complex lines---see the text that
follows~\eqref{eq:20}---so it is natural that partition functions of quantum
field theories are also elements of complex lines.)

\subsubsection*{Interlude: Remarks on the physics of projective theories}

We make two observations that may help locate our picture among more familiar
ideas in quantum field theory. 
 
The first is an interpretation of conversations with Nati Seiberg.  Namely, one
might consider two (linear) quantum field theories $F,F'$ to encode the same
physics if there is an invertible theory~$\lambda $ on the same background
fields such that\footnote{The tensor product operation on theories is often
called `stacking'.} $F'=F\otimes \lambda $.  After all, traditional correlation
functions are ratios, which in a path integral take the schematic form
  \begin{equation}\label{eq:36}
     \frac{\int_{}D\phi \;e^{-S(\phi )}\phi (x_1)\dots \phi
     (x_k)}{\int_{}D\phi \;e^{-S(\phi )}}, 
  \end{equation}
and such ratios are unchanged under tensoring by an invertible theory.  In
other terms, the categorical group of invertible field theories operates on the
collection of quantum field theories, and theories in the same orbit encode the
same physics.  Now simply observe that the orbits are projective
theories.\footnote{As stated here, they are projective theories with
trivializable anomaly.  We can modify the discussion to incorporate
nontrivializable anomalies by passing to centrally extended bordism
categories.}  Therefore, it is the projective theory that encodes physical
information.  
 
The second remark is about the following piece of common lore: \emph{A gapped
quantum system is well-approximated at low energy by a topological field
theory.}  In fact, the correct statement is that it is well-approximated by a
\emph{projective} field theory which is topological: a linear field theory
approximation may not be topological, but its projectivization is.  A showpiece
example that illustrates why projectivity is necessary is 3-dimensional
Yang-Mills with a nondegenerate Chern-Simons term.  Fix a compact Lie
group~$G$.  A general level is a cocycle that represents a class $\lambda \in
H^4(BG;\ZZ)$.  For $e\in \RR^{>0}$ consider the Wick-rotated theory with
lagrangian
  \begin{equation}\label{eq:37}
     L = \frac{1}{4e^2}F_A\wedge *F_A\;+\;\Gamma \mstrut _{\!\lambda }(A),
  \end{equation}
where $\Gamma \mstrut _{\!\lambda }(A)$~is the Chern-Simons term.  For this
3-dimensional theory, the background fields are 
  \begin{equation}\label{eq:38}
     \sF=\{\textnormal{orientations, Riemannian metrics}\}. 
  \end{equation}
The following claims are implicit in~\cite{W2}.

  \begin{claim}[]\label{thm:11}
 \ 
  \begin{enumerate}[label=\textnormal{(\arabic*)}]

 \item The lagrangian~\eqref{eq:37} determines a family of Wick-rotated field
theories parametrized by $e\in \RR^{>0}$.

 \item The singular limit as $e\to \infty $ exists and defines a field theory 
  \begin{equation}\label{eq:402}
     F\:\Btt(\sF)\longrightarrow \Vect. 
  \end{equation}

 \item The underlying \emph{projective} theory~$\bF$ is topological, i.e., it
factors through $\Btt(\sF')$ for 
  \begin{equation}\label{eq:403}
     \sF'=\{\textnormal{orientations}\} 
  \end{equation}
 
 \item The factored theory $\bF'\:\Btt(\sF')\to \Proj$ has a nontrivial
(framing) anomaly. 

 \end{enumerate}
  \end{claim}

\subsubsection*{Extension of an anomaly theory}

Often it happens that the anomaly theory~$\alpha $, which is a
once-categorified invertible $n$-dimensional theory, can be presented as the
truncation of an invertible $(n+1)$-dimensional theory~$\ta$: 
  \begin{equation}\label{eq:39}
     \begin{gathered}\xymatrix{ &  \BLine \\ 
      \Bord_{\langle n-1,n
     \rangle}(\sF)\ar@[red][ur]^<<<<<<<<<{\alpha}\ar@{^{(}->}[r] & 
      \Bord_{\langle n-1,n,n+1
     \rangle}(\widetilde{\sF})\ar@{-->}@[red][u]_{\tilde\alpha } } 
      \end{gathered} 
  \end{equation}
Here $\widetilde{\sF}$~is a sheaf of $(n+1)$-dimensional fields.  In this case,
a field theory~$F$ with anomaly~$\alpha $ can be realized as a boundary theory
of the invertible field theory~$\ta$.  We give an example of this at the end of
the paper: a free spinor field.

  \section{Anomalies as an obstruction to quantization}\label{sec:5}

In this section we explain the second half of Slogan~\eqref{eq:1}, which was
introduced at the start:
  $$ \textsc{Quantum theory is projective.  Quantization is linear.} $$
Fix a dimension~$n$, let $\sF$~be a sheaf of $n$-dimensional fields, and
suppose $\bF\:\Bn(\sF)\to \Proj$ is an $n$-dimensional (projective) theory
over~$\sF$.  We want to integrate out some fields in~$\sF$.  Hence assume that
$\sF$~is the total space of a fiber bundle
  \begin{equation}\label{eq:40}
     \pi \:\sF\longrightarrow  \sG 
  \end{equation}
of fields.  For the quantization of~$\bF$, we view the fibers of~$\pi $ as
\emph{fluctuating} fields and the base of~$\pi $ as \emph{background} fields.
A typical example is the fiber bundle
  \begin{equation}\label{eq:41} 
\begin{gathered}
     \xymatrix{\{\textnormal{orientations, Riemannian metrics,
     $G$-connections}\}\ar[d]^\pi \\ \{\textnormal{orientations, Riemannian
     metrics}\}}
     \end{gathered}
  \end{equation}
for $G$~a Lie group.  In this case we integrate over $G$-connections (gauge
fields) and treat orientations and Riemannian metrics as background fields.
Quantization of~$\bF$ along~$\pi $---integration over the fluctuating
fields---should produce a theory $\bG\:\Bn(\sG)\to \Proj$.  Of course, this
integration process has all the analytic interest and difficulties of quantum
field theory.  Our limited goal is to discuss descent of the projectivity
(anomaly) along~$\pi $.

  \begin{figure}[ht]
  \centering
  \includegraphics[scale=1.2]{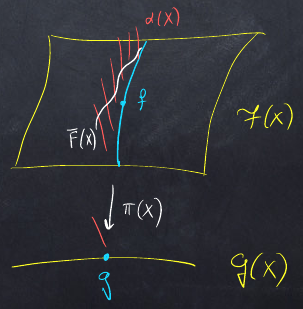}
  \vskip -.5pc
  \caption{The Feynman path integral on a closed $n$-manifold~$X$}\label{fig:2}
  \end{figure}

To see the issue, consider a closed $n$-manifold~$X$.  Integration over
fluctuating fields on~$X$ is depicted in \autoref{fig:2}.  Here $g\in \sG(X)$
is a background field on~$X$, and the fiber of~$\pi $ over~$g$ consists of
background + fluctuating fields $f\in \sF(X)$ that map to~$g$ under~$\pi $.
One must integrate the partition function of~$\bF$ over this fiber.  The
partition function~ $\bF(X)$ is a section of a line bundle $\alpha (X)\to
\sF(X)$.  One is instructed to integrate this section over the fibers of~$\pi
$.  However, that does not make formal sense: we cannot add elements of
different complex lines.\footnote{Of course, the integrand must be a density on
the fiber, not a function, but the important point is that it take values in a
\emph{fixed} line on each fiber, not a variable line.}  To carry out the
integration we need an isomorphism of the line bundle on each fiber to a
constant line bundle: the fiber over $g\in \sG(X)$ must be identified with a
fixed line that depends only on~$g$.  In other words, we must \emph{descend}
the line bundle $\alpha (X)\to \sF(X)$ to a line bundle over~$\sG(X)$.
(Descent is a parametrized trivialization along the fibers.)  This is the
argument for the anomaly as an obstruction to quantization that dates back to
the 1980's.  Also from that period is the variation of this argument for a
closed $(n-1)$-manifold~$Y$, in which case one is descending a gerbe.  This is
the \emph{Hamiltonian anomaly}~\cite{FS,NAg,Fa,S2}.

An extension of this discussion shows that we must descend the entire anomaly
theory~$\alpha $ to a once-categorified invertible theory over~$\sG$.
Trivialization of the anomaly on each fiber lifts from the projective to the
linear, which is the origin of the slogan~\eqref{eq:1}.  This descent has an
existence and uniqueness theory.  The existence question---anomaly as
obstruction---is the basis of the English word `anomaly' for the projectivity
of a field theory.
 
A descent~$\beta $ of the anomaly theory~$\alpha $ fits into the diagram 
  \begin{equation}\label{eq:42}
     \begin{gathered}
     \xymatrix@R-1pc@C+1.5pc{\Bn(\sF)\ar[dd]^\pi \ar@[red][dr]^\alpha \\ 
     &\BLine \\ \Bn(\sG)\ar@[red]@{-->}[ur]^<<<<<<<<{\beta }} \end{gathered} 
  \end{equation}
The data of the descent also includes an isomorphism 
  \begin{equation}\label{eq:43}
     \theta \:\alpha \xrightarrow{\;\;\cong \;\;}\pi ^*\beta 
  \end{equation}
A change of isomorphism~$\theta $ is multiplication by an invertible field
theory $\lambda \:\Bn(\sG)\to \Line$.

  \begin{remark}[]\label{thm:12}
 The descent of the anomaly is a tractable---and often very useful---piece of
data for the quantization of~$\bF$ along~$\pi $.  If one carries out that
quantization to produce a theory~$\bG$ over~$\sG$, then the anomaly theory
of~$\bG$ is \emph{not} necessarily~$\beta $: the quantization process may
introduce new contributions to the projectivity.  For example, if $\bF$~is a
classical (invertible) theory of a free spinor field---with a spin structure,
Riemannian metric, and possibly a connection as background fields---then the
anomaly~$\alpha $ is trivial and $\beta $~can be taken to be trivial as well.
But the quantum theory~$\bG$ of a free spinor field has a nontrivial anomaly. 
  \end{remark}

  \section{Anomaly of a spinor field}\label{sec:6}

As an illustration of anomalies, we review the general formula for the anomaly
of a free $n$-dimensional spinor field; this formula appears
in~\cite[\S9.2.5]{FH2}.  This anomaly theory extends to a full
$(n+1)$-dimensional invertible theory; as explained earlier, the anomaly is the
truncation to a once-categorified $n$-dimensional theory.  The formula for the
extended theory is a map of spectra in stable homotopy theory, and so we begin
with a quick review of the representation of invertible field theories as maps
of spectra.  Then we review the data that defines a free spinor field and
conclude with the formula~\eqref{eq:49} for the anomaly theory.  We remark that
the formula is a conjecture until more foundations (with differential
cohomology) are developed for invertible field theories and until the general
Wick-rotated free spinor field is completely constructed in the framework of
\autoref{thm:6}.

The account here is overly laconic, included mainly to illustrate
\autoref{thm:13}(1) below.  As compensation, we refer to~\cite{FH2} for
background, details, and references.

\subsubsection*{Invertible field theories and stable homotopy theory}

The basic idea is that an invertible theory 
  \begin{equation}\label{eq:44}
     \lambda \:\Bn(\sF)\longrightarrow \Line 
  \end{equation}
maps into a \emph{groupoid}: all morphisms in the category $\Line$ are
invertible.  Therefore, $\lambda $~factors through the \emph{quotient} of
$\Bn(\sF)$ in which all morphisms are inverted.  At that point we obtain a map
of groupoids, which is equivalently a map of spaces.  (For a leisurely
exposition of this equivalence, see~\cite{Po}.)  Furthermore, the symmetric
monoidal structure on the domain and codomain in~\eqref{eq:44} mean that the
spaces are equipped with an infinite loop structure.  Infinite loop spaces are
equivalent to (connective) spectra.  In the end, then, an invertible field
theory is represented as a map of spectra. 

  \begin{remark}[]\label{thm:13}
 \ 
 \begin{enumerate}[label=\textnormal{(\arabic*)}]

 \item The passage from categories to spectra makes invertible field theories
amenable to a suite of mathematical tools.

 \item The same passage applies to once-categorified invertible theories.
 
 \item Full locality is well-developed for \emph{topological} field theories,
and for invertible theories that are not necessarily topological we implicitly
assume full locality in what follows.  In the general invertible, possibly
non-topological, case we need \emph{differential} versions of spectra;
see~\cite{ADH} and references therein.
 \end{enumerate}
  \end{remark}

For a free spinor field we have background fields
  \begin{equation}\label{eq:45}
     \sF=\{\textnormal{spin structures, Riemannian metrics}\}. 
  \end{equation}
If we drop the Riemannian metrics, then the spectrum obtained from the
(extended) bordism category is the Thom spectrum~$\MSpin$.  In some cases the
anomaly theory we are after is topological, and then $\MSpin$~is the domain of
the anomaly theory.  In general, the domain is a differential version. 
 
There is a universal codomain for invertible field theories, characterized in
the topological case by the universal property that the partition functions
determine the theory.  We use a differential variant, which for $m$-dimensional
theories is a differential version of the \emph{Anderson dual to the sphere
spectrum} $\IZ$, shifted according to the dimension.  With differential
variants implicit, an $(n+1)$-dimensional invertible theory with background
fields~\eqref{eq:45} is a spectrum map
  \begin{equation}\label{eq:46}
     \alpha \:\MSpin\longrightarrow \Sigma ^{n+2}\IZ 
  \end{equation}

\subsubsection*{Spinor field data}

  \begin{figure}[ht]
  \centering
  \includegraphics[scale=1.35]{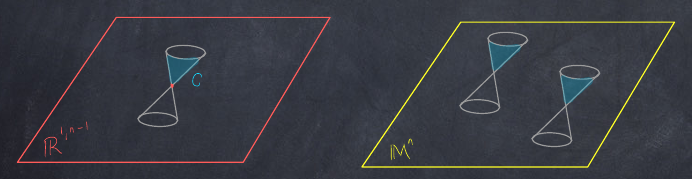}
  \vskip -.5pc
  \caption{Minkowski spacetime~$\MM^n$ and the vector space
  $\RR^{1,n-1}$}\label{fig:3} 
  \end{figure}

This data is given in the relativistic setting.  Fix a spacetime dimension~$n$.
Let $\MM^n$~denote standard Minkowski spacetime.  It is an $n$-dimensional
affine space acted upon simply transitively by the vector group~$\RR^{1,n-1}$
of translations; see \autoref{fig:3}.  $\MM^n$~ is equipped with a
translation-invariant Lorentz metric and a choice $C\subset \RR^{1,n-1}$ of a
component of (forward) timelike vectors---a time-orientation.  The spin group
$\Spin_{1,n-1}\subset \Cliff^0_{n-1,1}$ is a subset of the even Clifford
algebra.\footnote{This Clifford algebra has $n-1$ generators that square
to~$+1$ and one generator that squares to~$-1$.}  Spinor field data is a
triple~$(\SS,\Gamma ,m)$ that consists of
  \begin{equation}\label{eq:47}
     \begin{alignedat}{2} &{ \SS }&&\textnormal{real (ungraded) finite
     dimensional 
      $\Cliff^0_{n-1,1}$-module} \\ &{ \Gamma \:\SS\times
      \SS\longrightarrow \RR^{1,n-1}}\qquad \;&&\textnormal{symmetric
      $\Spin_{1,n-1}$-invariant form; $\Gamma (s,s)\in \overline{C}$ for all
      $s\in \SS$}\\ &{ m\:\SS\times \SS\longrightarrow
      \RR}&&\textnormal{skew-symmetric $\Spin_{1,n-1}$-invariant
      (\emph{mass}) form}\end{alignedat} 
  \end{equation}
The Clifford module~$\SS$ restricts to a spin representation
of~$\Spin_{1,n-1}$.  It is a remarkable theorem that the positive symmetric
pairing~$\Gamma $ exists for all~$\SS$; this is only true in Lorentz signature.
Furthermore, if $\SS$~is irreducible, then $\Gamma $~is unique up to positive
scale; in general there is a contractible space of possible~$\Gamma $.  The
mass form~$m$ may vanish.
 
Several important algebraic facts are proved or referenced
in~\cite[\S9.2.4]{FH2}.  Given $(\SS,\Gamma )$, there is a unique compatible
$\zt$-graded $\Cliff_{n-1,1}$-module structure on~$\SS\oplus \SS^*$.
Furthermore, every finite dimensional $\Cliff_{n-1,1}$-module has this form.
Crucially, nondegenerate mass forms for~$(\SS,\Gamma )$ correspond to
$\Cliff_{n-1,2}$-module structures on $\SS\oplus \SS^*$ that extend the
$\Cliff_{n-1,1}$-module structure.

\subsubsection*{The anomaly theory}

We give the conjectured formula for~$m=0$.  In~\cite{CFLS} the theory with
$m$~as a scalar field is considered, and there is a conjectured formula for the
anomaly in~\cite[\S7.4]{CFLS}.
 
Since $\Gamma $~is a contractible choice, we drop it from the notation.
Atiyah-Bott-Shapiro~\cite{ABS} identify the abelian group of equivalence
classes of $\Cliff_{n-1,1}$-modules modulo those that extend to
$\Cliff_{n-1,2}$-modules in terms of the $KO$-theory spectrum, and the
module~$\SS$ determines a class in that group: 
  \begin{equation}\label{eq:48}
     [\SS ]\in \pi _{2-n}KO\cong [S^0,\Sigma ^{n-2}KO] ,
  \end{equation}
where $S^0$~is the sphere spectrum.   

  \begin{claim}[]\label{thm:15}
 The anomaly theory~$\alpha = \aSG$ of the massless free spinor field is a
differential lift of 
  \begin{equation}\label{eq:49}
     \MSpin\xrightarrow{\;\;\phi \wedge [\SS ]\;\;}KO\wedge \Sigma
     ^{n-2}KO\xrightarrow{\;\;\mu \;\;}\Sigma
     ^{n-2}KO\xrightarrow{\;\;\Pfaff\;\;}\Sigma ^{n+2}\IZ 
  \end{equation}
  \end{claim}

\noindent
 Here $\phi \:\MSpin\to KO$ is essentially the Thom class of a real spin bundle,
also defined in~\cite{ABS}.  The map~$\mu $ is multiplication in the ring
spectrum~$KO$, and $\Pfaff\:KO\to \Sigma ^4\IZ$ is the map that enters the
Anderson self-duality of~$KO$; see~\cite{FMS}. 

  \begin{remark}[]\label{thm:16}
 As written (i.e., without a differential lift), \eqref{eq:49}~is the
\emph{deformation class} of the anomaly theory~$\alpha $. 
  \end{remark}

\appendix

   \section{On the definition of $\Proj$}\label{sec:7}

We make some comments on the definition of~$\Proj$, introduced in~\eqref{eq:27}
as the codomain of a projective field theory~\eqref{eq:33}.  Our treatment is
sparse since a detailed development is the subject of a forthcoming thesis of
Chetan Vuppulury (a student of Domenico Fiorenza). 
 
Warmup: Suppose $S$~is a set equipped with the action of a group~$G$.  The
naive quotient is the set~$S/G$ of orbits of the $G$-action.  But it flouts the
core ethos of categorical thinking to identify elements of~$S$ to form a
quotient.  Rather, one should remember the group element that effects the
identification.  This leads to the action groupoid~$S\gpd G$ whose set of
objects is~$S$ and whose set of morphisms is $S\times G$.  Lesson: The quotient
of a set by a group is a 1-category. 
 
The sequence~\eqref{eq:27} exhibits $\Proj$ as the quotient of $\Vect$ by
$\Line$, and the warmup suggests that this quotient is a 2-category.  A model
for $\Line$ is the groupoid $B\Cx$; it has a single object with automorphism
group~$\Cx$.  Hence the 2-category $\Proj$ has vector spaces as objects,
linear maps as 1-morphisms, and diagrams 
  \begin{equation}\label{eq:50}
     \begin{gathered} \xymatrix@C+2pc{V\rtwocell<5> ^{T_0}_{T_1}{\lambda } & V'}
     \end{gathered} 
  \end{equation}
as 2-morphisms, where $\lambda \in \Cx$ and the linear maps $T_0,T_1$ satisfy
$T_1=\lambda T_0$.  (The same adjectives used to define $\Vect$---certain
\emph{topological} vector spaces, \emph{nuclear} linear maps---apply to $\Proj$.)
 
More generally, suppose $\sC$ is a symmetric monoidal $(\infty ,n+1)$-category
for some\footnote{The construction also works for a symmetric monoidal $(\infty
,0)$-category, represented as an infinite loop space~$X$.  Then $\PP\Omega
X$~is the space of paths in~$X$ emanating from the basepoint.} $n\in
\ZZ^{\ge0}$.  Let $\sCx\subset \sC$ be the maximal Picard subgroupoid of~$\sC$.
Heuristically, define the \emph{projectivization} $\POC$ of $\Omega
\sC=\Hom_{\sC}(1,1)$ as the pullback
  \begin{equation}\label{eq:51}
     \begin{gathered} \xymatrix{\POC\ar@{-->}[r]^{} \ar@{-->}[d]_{} &
     \sCx\ar[d]^{} \\ \ast\ar[r]^{} & \sC} \end{gathered} 
  \end{equation}
The special case that defines $\Proj$ is 
  \begin{equation}\label{eq:52}
     \begin{gathered} \xymatrix{\Proj\ar@{-->}[r]^{} \ar@{-->}[d]_{} &
     B\Line\ar[d]^{} \\ \ast\ar[r]^{} & \Alg} \end{gathered} 
  \end{equation}
where $\Alg$ is the Morita 2-category of algebras, bimodules, and intertwiners.
The core issue is to \emph{define} the pullback~\eqref{eq:51}.  In fact, more
general constructions are given in~\cite{JS}.  Namely, Johnson-Freyd and
Scheimbauer define categories $\sC^{\downarrow}$ and $\sC^{\rightarrow}$, each
equipped with source and target homomorphisms $s,t\:\sC^*\to \sC$, where
$*=\downarrow$ or $*=\rightarrow$.  The pullback we seek is presumably the
intersection $s\inv (1)\cap t\inv (\sCx)$.  Finally, in terms of
\cite[Definition~1.4]{JS} and the papers that inspired it, we see that a
projective field theory is a field theory \emph{relative} to its anomaly
theory.

 \bigskip\bigskip
\providecommand{\bysame}{\leavevmode\hbox to3em{\hrulefill}\thinspace}
\providecommand{\MR}{\relax\ifhmode\unskip\space\fi MR }
\providecommand{\MRhref}[2]{%
  \href{http://www.ams.org/mathscinet-getitem?mr=#1}{#2}
}
\providecommand{\href}[2]{#2}


\begin{thebibliography}{AGVM22}

\bibitem[A]{A}
M.~F. Atiyah, \emph{Topological quantum field theories}, Inst. Hautes
  {\'E}tudes Sci. Publ. Math. (1988), no.~68, 175--186 (1989).

\bibitem[ABS]{ABS}
M.~F. Atiyah, R.~Bott, and A.~A. Shapiro, \emph{Clifford modules}, Topology
  \textbf{3} (1964), 3--38.

\bibitem[Ad]{Ad}
Stephen~L Adler, \emph{Axial-vector vertex in spinor electrodynamics}, Physical
  Review \textbf{177} (1969), no.~5, 2426.

\bibitem[ADH]{ADH}
Araminta Amabel, Arun Debray, and Peter~J. Haine, \emph{Differential
  Cohomology: Categories, Characteristic Classes, and Connections},
  \href{http://arxiv.org/abs/arXiv:2109.12250}{{\tt arXiv:2109.12250}}.

\bibitem[AgVm]{AgVm}
Luis Alvarez-Gaume and Miguel~A. Vazquez-Mozo, \emph{{Anomalies and the
  Green-Schwarz Mechanism}}, \href{http://arxiv.org/abs/2211.06467}{{\tt
  arXiv:2211.06467 [hep-th]}}.

\bibitem[BH]{BH}
Dorje~C. Brody and Lane~P. Hughston, \emph{Geometric quantum mechanics},
  \href{http://dx.doi.org/10.1016/S0393-0440(00)00052-8}{J. Geom. Phys.
  \textbf{38} (2001)}, no.~1, 19--53.

\bibitem[BJ]{BJ}
John~Stewart Bell and Roman~W Jackiw, \emph{A PCAC puzzle: $\pi^0 \to\gamma
  \gamma$ in the $\sigma $-model}, Nuovo cimento \textbf{60} (1969),
  no.~CERN-TH-920, 47--61.

\bibitem[Bo]{Bo}
L.~{Bonolis}, \emph{{From the Rise of the Group Concept to the Stormy Onset of
  Group Theory in the New Quantum Mechanics. A saga of the invariant
  characterization of physical objects, events and theories.}},
  \href{http://dx.doi.org/10.1393/ncr/i2004-10006-4}{Nuovo Cimento Rivista
  Serie \textbf{27} (2004)}, no.~4, 040000--110.

\bibitem[CFLS]{CFLS}
Clay C\'ordova, Daniel~S. Freed, Ho~Tat Lam, and Nathan Seiberg,
  \emph{{Anomalies in the Space of Coupling Constants and Their Dynamical
  Applications I}},
  \href{http://dx.doi.org/10.21468/SciPostPhys.8.1.001}{SciPost Phys.
  \textbf{8} (2020)}, no.~1, 001, \href{http://arxiv.org/abs/1905.09315}{{\tt
  arXiv:1905.09315 [hep-th]}}.

\bibitem[CH]{CH}
Curtis~G. Callan, Jr. and Jeffrey~A. Harvey, \emph{{Anomalies and Fermion Zero
  Modes on Strings and Domain Walls}},
  \href{http://dx.doi.org/10.1016/0550-3213(85)90489-4}{Nucl. Phys.
  \textbf{B250} (1985)},
427--436.

\bibitem[F1]{F1}
Daniel~S Freed, \emph{Anomalies and invertible field theories}, Proc. Symp.
  Pure Math, Proc. Sympos. Pure Math., vol.~88, Amer. Math. Soc., Providence,
  RI, 2014, pp.~25--45. \href{http://arxiv.org/abs/arXiv:1404.7224}{{\tt
  arXiv:1404.7224}}.

\bibitem[F2]{F2}
Daniel~S. Freed, \emph{On Wigner's theorem}, Proceedings of the Freedman Fest
  (Vyacheslav~Krushkal Rob~Kirby and Zhenghan Wang, eds.), Geometry \& Topology
  Monographs, vol.~18, Mathematical Sciences Publishers, 2012, pp.~83--89.
  \href{http://arxiv.org/abs/arXiv:1211.2133}{{\tt arXiv:1211.2133}}.

\bibitem[Fa]{Fa}
L.~D. Faddeev, \emph{Hamiltonian approach to the theory of anomalies}, Recent
  Developments in Mathematical physics (H.~Mitter and L.~Pittner, eds.),
  Internationale Universitatswoche fur Kernphysik, Schladming, Austria,
  vol.~26, 1987, pp.~137--159.

\bibitem[FH1]{FH1}
Daniel~S. Freed and Michael~J. Hopkins, \emph{Chern-{W}eil forms and abstract
  homotopy theory},
  \href{http://dx.doi.org/10.1090/S0273-0979-2013-01415-0}{Bull. Amer. Math.
  Soc. (N.S.) \textbf{50} (2013)}, no.~3, 431--468,
  \href{http://arxiv.org/abs/arXiv:1301.5959}{{\tt arXiv:1301.5959}}.

\bibitem[FH2]{FH2}
\bysame, \emph{Reflection positivity and invertible topological phases},
  \href{http://dx.doi.org/10.2140/gt.2021.25.1165}{Geom. Topol. \textbf{25}
  (2021)}, no.~3, 1165--1330, \href{http://arxiv.org/abs/arXiv:1604.06527}{{\tt
  arXiv:1604.06527}}.

\bibitem[FHT]{FHT}
Daniel~S. Freed, Michael~J. Hopkins, and Constantin Teleman,
  \href{http://dx.doi.org/10.1093/acprof:oso/9780199534920.003.0019}{\emph{Consistent
  orientation of moduli spaces}}, The many facets of geometry, Oxford Univ.
  Press, Oxford, 2010, pp.~395--419.
  \href{http://arxiv.org/abs/arXiv:0711.1909}{{\tt arXiv:0711.1909}}.

\bibitem[FMS]{FMS}
Daniel~S. Freed, Gregory~W. Moore, and Graeme Segal, \emph{The uncertainty of
  fluxes}, \href{http://dx.doi.org/10.1007/s00220-006-0181-3}{Commun. Math.
  Phys. \textbf{271} (2007)}, 247--274,
\href{http://arxiv.org/abs/hep-th/0605198}{{\tt arXiv:hep-th/0605198}}.

\bibitem[FS]{FS}
L.D. Faddeev and S.~L. Shatashvili, \emph{{Algebraic and Hamiltonian Methods in
  the Theory of Nonabelian Anomalies}},
  \href{http://dx.doi.org/10.1007/BF01018976}{Theor. Math. Phys. \textbf{60}
  (1985)}, 770--778.

\bibitem[FT]{FT}
Daniel~S. Freed and Constantin Teleman, \emph{Relative quantum field theory},
  \href{http://dx.doi.org/10.1007/s00220-013-1880-1}{Comm. Math. Phys.
  \textbf{326} (2014)}, no.~2, 459--476,
  \href{http://arxiv.org/abs/arXiv:1212.1692}{{\tt arXiv:1212.1692}}.

\bibitem[JS]{JS}
Theo Johnson-Freyd and Claudia Scheimbauer, \emph{(Op)lax natural
  transformations, twisted quantum field theories, and ``even higher'' Morita
  categories}, Advances in Mathematics \textbf{307} (2017), 147--223,
  \href{http://arxiv.org/abs/arXiv:1502.06526}{{\tt arXiv:1502.06526}}.

\bibitem[KS]{KS}
Maxim Kontsevich and Graeme Segal, \emph{Wick rotation and the positivity of
  energy in quantum field theory},
  \href{http://dx.doi.org/10.1093/qmath/haab027}{Q. J. Math. \textbf{72}
  (2021)}, no.~1-2, 673--699, \href{http://arxiv.org/abs/arXiv:2105.10161}{{\tt
  arXiv:2105.10161}}.

\bibitem[La]{La}
R.~Lashof, \emph{Poincar\'e duality and cobordism}, Trans. Amer. Math. Soc.
  \textbf{109} (1963), 257--277.

\bibitem[Lo]{Lo}
Ottmar Loos, \emph{Symmetric spaces. {II}: {C}ompact spaces and
  classification}, W. A. Benjamin, Inc., New York-Amsterdam, 1969.

\bibitem[M]{M}
John~W. Milnor, \emph{Lectures on the {$h$}-cobordism theorem}, Notes by L.
  Siebenmann and J. Sondow, Princeton University Press, Princeton, N.J., 1965.

\bibitem[NAg]{NAg}
Philip Nelson and Luis Alvarez-Gaum{{\'e}}, \emph{Hamiltonian interpretation of
  anomalies}, Comm. Math. Phys. \textbf{99} (1985), no.~1, 103--114.

\bibitem[Po]{Po}
Timothy Porter, \emph{Spaces as infinity-groupoids}.
  \url{https://ncatlab.org/nlab/files/Spaces+as+infinity-groupoids.pdf}.

\bibitem[S1]{S1}
Graeme Segal, \emph{The definition of conformal field theory}, Topology,
  geometry and quantum field theory, London Math. Soc. Lecture Note Ser., vol.
  308, Cambridge Univ. Press, Cambridge, 2004, pp.~421--577.

\bibitem[S2]{S2}
G.~B. Segal, \emph{Faddeev's anomaly in Gauss's law}. preprint.

\bibitem[St]{St}
J~Steinberger, \emph{On the use of subtraction fields and the lifetimes of some
  types of meson decay}, Physical Review \textbf{76} (1949), no.~8, 1180.

\bibitem[tH]{tH}
Gerard 't~Hooft, \emph{{Naturalness, chiral symmetry, and spontaneous chiral
  symmetry breaking}},
  \href{http://dx.doi.org/10.1007/978-1-4684-7571-5_9}{NATO Sci. Ser. B
  \textbf{59} (1980)},
135--157.

\bibitem[vNW]{vNW}
J.~v.~Neumann and E.~Wigner, \emph{Zur Erkl{\"a}rung einiger Eigenschaften der
  Spektren aus der Quantenmechanik des Drehelektrons},
  \href{http://dx.doi.org/10.1007/BF01327823}{Zeitschrift f{\"u}r Physik A
  Hadrons and Nuclei \textbf{49} (1928)}, 73--94.

\bibitem[W1]{W1}
Edward Witten, \emph{{World sheet corrections via D instantons}},
  \href{http://dx.doi.org/10.1088/1126-6708/2000/02/030}{JHEP \textbf{02}
  (2000)}, 030,
\href{http://arxiv.org/abs/hep-th/9907041}{{\tt arXiv:hep-th/9907041}}.

\bibitem[W2]{W2}
\bysame, \emph{Quantum field theory and the {J}ones polynomial}, Comm. Math.
  Phys. \textbf{121} (1989), no.~3, 351--399.

\end{thebibliography}
  \end{document}